\documentclass[usenatbib]{mn2e}

\usepackage{psfig,morefloats,url}

\advance \voffset by -1.5cm\relax

\def\lapp{\ifmmode\stackrel{<}{_{\sim}}\else$\stackrel{<}{_{\sim}}$\fi}

\footnotesize
\newdimen\digitwidth    %define ! a one digit width for tables
\setbox0=\hbox{\rm0}
\digitwidth=\wd0
\catcode`!=\active
\def!{\kern\digitwidth}
\normalsize
\def\msun{\,{\rm M}_{\odot}}
\def\ltsim{\ifmmode\stackrel{<}{_{\sim}}\else$\stackrel{<}{_{\sim}}$\fi}
\def\gtrsim{\ifmmode\stackrel{>}{_{\sim}}\else$\stackrel{>}{_{\sim}}$\fi}

\title[Double and single recycled pulsars]
{Double and single recycled pulsars: an evolutionary puzzle?}
\author[Belczynski et al.] 
{
K.~Belczynski$^{1,2,3}$\thanks{Email:kbelczyn@nmsu.edu}, 
D.R.~Lorimer$^{4,5}$,
J.P.~Ridley$^4$ and S.J.~Curran$^{6}$\\ 
$^{1}$ Los Alamos National Laboratory, New Mexico, USA\\
$^{2}$ Astronomical Observatory, University of Warsaw, Al. Ujazdowskie 4,
            00-478 Warsaw, Poland\\
$^{3}$ Oppenheimer Fellow\\
$^{4}$ Department of Physics, West Virginia University, Morgantown, 
       WV 26506, USA\\
$^{5}$ Green Bank Observatory, PO Box 2, Green Bank, WV 24944, USA\\
$^{6}$ School of Physics, University of New South Wales, Sydney, NSW 2052, 
Australia
}
\let\leqslant=\leq %Authors with AMS fonts and mssymb.tex can comment
\date{\today}
\begin{document}

\maketitle
\newcommand{\setthebls}{
%                 de-comment this line for double spacing:
 %\baselineskip=20pt
}

\setthebls

\begin{abstract}
We investigate the statistics of isolated recycled pulsars and double
neutron star binaries in the Galactic disk. Since recycled pulsars are
believed to form through accretion and spinup in close binaries, the
isolated objects presumably originate from disrupted progenitors of double
neutron stars.  There are a comparable number of
double neutron star systems compared to isolated recycled pulsars.
We find that standard evolutionary models cannot explain
this fact, predicting several times the number of
isolated recycled pulsars than those in double neutron star systems.
We demonstrate, through population synthesis calculations,
that the velocity distribution of isolated recycled pulsars is broader
than for binary systems. When this is accounted for in a model for
radio pulsar survey selection effects, which include the effects of
Doppler smearing for the double neutron star binaries, we find that
there is a small ($\sim 25\%$) bias towards the detection of double 
neutron star systems. This bias, however, is not significant enough
to explain the observational discrepancy if standard ($\sigma = 265$~km~s$^{-1}$)
neutron star natal kick velocities are invoked in binary population syntheses.
Population syntheses in which the 1D Maxwellian velocity dispersion of the
natal kick is $\sigma \sim 170$~km~s$^{-1}$ are consistent with 
the observations. These conclusions further support earlier findings 
the neutron stars formed in close interacting binaries receive 
significantly smaller natal kicks than the velocities of Galactic single 
pulsars would seem to indicate.
\end{abstract}

\begin{keywords}
methods: statistical; stars: neutron; stars: kinematics; pulsars: general
\end{keywords}

\section{Introduction}\label{sec:intro}

The pulsar population is an important tool to study various aspects of
stellar evolution, supernova rates, birth properties of neutron stars
and the evolution of massive binary systems.  Of particular interest
are the double neutron star (DNS) binary systems whose inspirals are
one of the key events expected by the gravitational wave detectors now
in operation \citep[see, e.g.,][]{aaa+06}. DNS binaries consist of an
older neutron star with a short spin period (typically in the range
20--100 ms) formed in the supernova explosion of the initially more
massive star in the binary system (the primary). The first-born
neutron star initially behaves as a regular radio pulsar, but
subsequently becomes spun up (recycled) via the accretion of matter
during Roche lobe overflow (RLOF) from the secondary star once it
leaves the main sequence. Following the supernova explosion of the
secondary, the resulting DNS consists of a recycled pulsar and a
younger second-born neutron star.

While a number of studies have addressed the population, lifetime and
merger rate of DNS binaries \citep[e.g.,][]{phi92,kkl03,cb05,ikb06},
less attention has been given to those binary systems which disrupt
during the second supernova \citep[see, however,][]{kl00}.  Of
particular significance are the statistics of the recycled pulsars
released from these binary systems, with their distinct spin
properties.  As discussed by \citet{lma+04}, these so-called
`disrupted recycled pulsars' (DRPs) directly probe the fraction of DNS
binaries which do not survive the second supernova explosion and can
therefore provide an independent constraint on population synthesis
models which predict a certain fraction of DRPs relative to DNS
binaries.

As pointed out by \citet{lma+04}, there is an apparent conundrum posed
by the observed DNS/DRP statistics. Given our current understanding of
binary evolution, the ratio of the {\it underlying} number of DNS
systems to DRPs observed should equal the survival fraction
for binary systems during the second supernova event, i.e.~those which
remained bound after the explosion. \citet{lma+04} estimated this fraction
to be around 0.1 based on scale-factor analysis of DNSs and DRPs and appeared
to be in reasonable accord with theoretical estimates of the survival
fraction taken
from the literature \citep{py98}.  Assuming
that the luminosities and radio lifetimes of the recycled pulsars
observable in DNS binaries are identical to the DRPs, the above
estimate implies that we should see roughly ten DRPs for every DNS
binary. However, this expectation is not confirmed in the observed
sample discussed in Section~\ref{sec:sample} where there are
comparable numbers of DRPs and DNS binaries. 

In light of these issues, the relationship between DNS binaries and
DRPs is an interesting problem which we address in this paper using
Monte Carlo simulations of binary populations and observational
selection effects. The goal of this work is to understand the
relationship between the observed and underlying ratios of
DNS binaries to DRPs. This can be summarized by the expression
\begin{equation}
  r_{\rm obs} = r_{\rm int} f_{\rm obs},
\end{equation}
where $r_{\rm obs}$ is the observed ratio of DNS binaries to DRPs,
$r_{\rm int}$ is the underlying (intrinsic) ratio and $f_{\rm obs}$ is a
correction factor which takes account of observational selection
effects.  As we discuss in Section~\ref{sec:sample},
we find that $r_{\rm obs} \sim 1$. In Sections
\ref{sec:simulation} and 4, we use state-of-the-art
binary population synthesis models to explore the possible predicted
ranges of $r$.  We investigate observational selection
effects in radio pulsar surveys to evaluate $f_{\rm obs}$ in Section
\ref{sec:selfx}.  Finally, in Section \ref{sec:conclusions}, we
summarize the main findings of this study.

\begin{table}
\caption{\label{tab:dnsdrp}
Spin and spatial properties of DNS binaries and DRPs currently known in
the Galactic
disk. From left to right, the columns list pulsar name, spin period
($P$ in ms), the base-10 logarithms of characteristic age ($\tau =
P/(2 \dot{P})$ in yr) and inferred magnetic field strength ($B \propto
(P\dot{P})^{1/2}$ in Gauss), distance ($d$ in kpc) and height above
the Galactic plane ($z$ in pc). The latter two quantities are based on
the pulsar dispersion measure and the Cordes \& Lazio (2002) model for
the Galactic distribution of free electrons.  The right-hand column
lists the discovery paper for each pulsar.
}
\label{tb:binary}
\begin{tabular}{lrrrrrr}
\hline
PSR & 
\multicolumn{1}{c}{$P$} & 
\multicolumn{1}{c}{log $\tau$} & 
\multicolumn{1}{c}{log $B$} & 
\multicolumn{1}{c}{$d$} & 
\multicolumn{1}{c}{$|z|$} &
\multicolumn{1}{c}{Ref.$^1$} \\
%    & 
%\multicolumn{1}{c}{(ms)}&  
%\multicolumn{1}{c}{log (yr)}& 
%\multicolumn{1}{c}{log (G)} & 
%\multicolumn{1}{c}{(kpc)} & 
%\multicolumn{1}{c}{(pc)} &  \\
\hline
\multicolumn{7}{c}{Compact DNS systems (binary period $<$ 1~day)} \\
\hline
J0737$-$3039 & 22.7& 8.3 & 9.8 & 0.6  & 40 & 1 \\
B1534$+$12   & 37.9& 9.4 & 10.0& 0.9  & 670& 2 \\
J1756$-$2251 & 28.5& 8.6 & 9.7 & 2.9  & 50 & 3 \\
B1913$+$16   & 59.0& 8.0 & 10.4& 7.1  & 260& 4 \\
\hline
\multicolumn{7}{c}{Wide DNS systems (binary period $>$ 1~day)} \\
\hline
J1518$+$4904 & 40.9& 10.4& 9.0 & 0.7  & 570& 5 \\
J1753$-$2240 & 95.1&  9.2&10.0 & 3.0  &  90& 6 \\
J1811$-$1736 &104.2&  9.0&10.1 & 5.9  &  50& 7 \\
J1829$+$2456 & 41.0& 10.1& 9.2 & 0.8  & 200& 8 \\
\hline
% P age B D |z|
\multicolumn{7}{c}{DRPs ($B<3.0 \times 10^{10}$~G)$^2$} \\
\hline
J0609$+$2130 & 55.7& 9.6 & 9.6 & 1.8  &  30& 9  \\
J1038$+$0032 & 28.9& 9.8 & 9.1 & 2.4  &1800& 10 \\
J1320$-$3512 &458.5& 9.6 &10.5 & 0.9  & 430& 11 \\
J1333$-$4449 &345.6&10.0 &10.1 & 2.3  & 690& 12 \\
J1339$-$4712 &137.1& 9.6 & 9.9 & 1.8  & 450& 12 \\
J1355$-$6206 &276.6& 9.1 &10.5 & 8.0  &  20& 13 \\
J1548$-$4821 &145.7& 9.5 &10.0 & 3.8  & 310& 13 \\
J1611$-$5847 &354.6& 9.4 &10.4 & 2.4  & 230& 14 \\
J1753$-$1914 & 63.0& 8.7 &10.1 & 2.7  & 160& 14 \\
J1816$-$5643 &217.9& 9.2 &10.3 & 3.1  & 940& 12 \\
B1952$+$29   &426.7& 9.6 &10.4 & 0.4  &  10& 15 \\
J2235$+$1506 & 59.8& 9.8 & 9.5 & 1.2  & 680& 16 \\
\hline
\end{tabular}

$^{1}$ -- The references used in this compilation are
 1: \cite{bdp+03}, 
 2: \cite{wol91a}, 
 3: \cite{fkl+05}, 
 4: \cite{ht75a}, 
 5: \cite{nst96}, 
 6: \cite{kkl+09},
 7: \cite{lcm+00}, 
 8: \cite{clm+04}, 
 9: \cite{lma+04}, 
10: \cite{bjd+06}, 
11: \cite{mld+96},
12: \cite{jbo+07},
13: \cite{kbm+03},
14: \cite{lfl+06}, 
15: \cite{dlp70},
16: \cite{cnt93}.\\
$^{2}$ -- This population of 12 single NSs may be contaminated by 
$\sim 4$ regular (non recycled) NSs, and therefore the number of 
known DRPs (binary origin) is $\sim 8$ (see Sec.2.2). 

\end{table}

\begin{table}
\caption{
Results of population synthesis calculations for the DNS and DRP populations. 
From left to right, the columns give the name of evolutionary model,
natal kick velocity dispersion, 
intrinsic numbers of DNS ($n_{\rm DNS}$) and DRPs ($n_{\rm DRP}$), and the 
intrinsic ratio $r = n_{\rm DNS}/n_{\rm DRP}$. The range of values
corresponds to changing evolutionary assumption on common envelope
evolution. 
}
\label{popsyn}
\begin{tabular}{lcccc}
\hline
Model & $\sigma_{\rm CC}$ & $n_{\rm DNS}$ & $n_{\rm DRP}$ & $r$ \\
      &   km~s$^{-1}$     &               &               &     \\
\hline
 A & 265
   &  3854--2977  & 13418--10085  & 0.29--0.30 \\
 B & 199
   &  5798--4295  & 11166--8316  & 0.52--0.53 \\
 C & 133
   & 9587--7252  &  8476--5911  & 1.13--1.23 \\
 D &  66
   & 18042--13572 &  8141--3166  & 2.22--4.36 \\
 E &   0
   & 42145--36316 & 46817--15263 & 0.90--2.38 \\
\hline
\end{tabular}
\end{table}

\section{The observational sample}\label{sec:sample}

\begin{figure}
\psfig{file=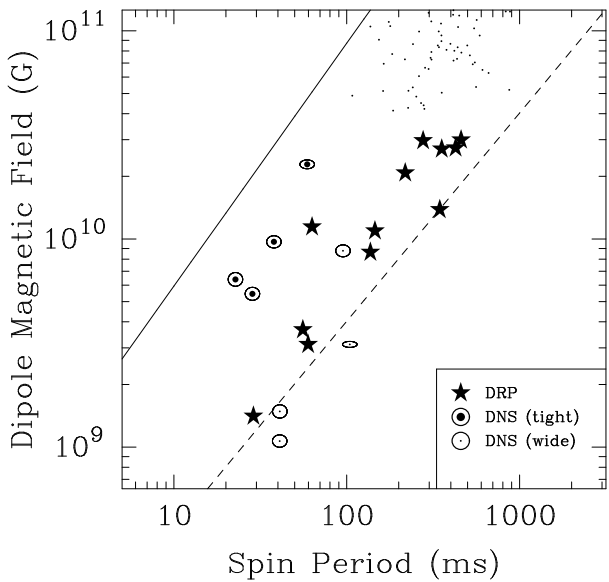,width=8cm,angle=0}
\caption{\label{fig:bp}
Magnetic field--period ($B-P$) diagram showing the samples of DNS
binaries, DRPs and other isolated radio pulsars. The DNS systems
are highlighted by ellipses with the eccentricity of the ellipse
representing the orbital eccentricity. Compact DNS binary systems
which will merge within a Hubble time are marked with larger dots.
DRPs, defined as isolated pulsars with $B<3 \times 10^{10}$~G and $P>20$~ms,
are shown as filled stars. The solid line is the limiting spin-up
period for Alfv\`en accretion at the Eddington limit. 
The dashed
line is the locus of points with characteristic age equal to 10~Gyr.
}
\end{figure}

Table~\ref{tab:dnsdrp} summarizes the observational data for the known
DNS binaries and DRPs in the Galactic disk. In Fig.\ref{fig:bp},
we present an updated version of the magnetic field--period ($B-P$) diagram
from Lorimer et al.~(2004) showing both samples of objects. 
There are currently nine DNS binaries which can be identified based on
their orbital parameters where measurements of multiple post-Keplerian
parameters (see, e.g.~Lorimer 2008) suggest the presence of two
neutron stars in each system. For the purposes of this paper, where the
focus is on recycled pulsars produced during binary evolution, we do not
select PSR~J1906+0746~\citep{lsf+06} where the observed radio pulsar is
likely the young second-born neutron star formed in the binary. The resulting
sample therefore consists of eight objects. We define a DRP as an isolated 
pulsar in the Galactic disk with $B<3 \times 10^{10}$~G and $P>20$~ms.
The latter criterion ensures that no isolated millisecond pulsars, which
are thought to have had a different evolutionary history, are included 
in our sample. The actual value of the limiting spin period was chosen
such that recycled pulsars in the known DNS sample would have been selected 
had their hosting binaries been disrupted. The isolated millisecond pulsars 
with $B<3 \times 10^{10}$~G and $P<20$~ms (about 28 known) are believed to 
accreted from a low mass companion (e.g., a white dwarf) over long period 
of time ($\sim 10^8$ yr) and then the companion was evaporated (e.g., 
Lorimer et al. 2004 and references therein). The population of isolated 
pulsars with small magnetic field ($B<3 \times 10^{10}$~G) but larger spin 
periods ($P>20$~ms) is believed to accreted from a high mass companion over 
relatively short period of time ($\sim 10^{6-7}$ yr) and then the 
companion was ejected from a binary while undergoing supernova explosion. 
There are a dozen such objects. We now discuss possible selection biases 
present in these samples, and draw some simple conclusions based on the 
available data.

\subsection{The DNS sample}

In Table~\ref{tab:dnsdrp}, we 
subdivide the DNS sample into ``compact systems'' with orbital periods
less than one day which will merge due to gravitational wave emission
within a Hubble time and ``wide systems'' with longer orbital periods
that will effectively never merge on relevant timescales. Despite the
small-number statistics, it is apparent, both from Table~1 and Fig.~1
that the compact systems appear to be younger than the wide
systems. As noted by several previous authors \citep{phi92,acw99,cb05},
this is a selection effect caused by the shorter coalescence time
compared to the radio lifetimes of these systems. The wide DNS
binaries and DRPs effectively spin-down until they reach the so-called
``death line'' at which radio emission is thought to become
ineffective and cease for all radio pulsars \cite{cr93}. In our
simulations of the underlying and observed DNS sample described below,
we will account for this important selection effect by carefully 
modeling both the orbital evolution (Section 3) and Doppler smearing
(Section 4) of such systems.

\subsection{The DRP sample}

Our choice of a maximum magnetic field of $3 \times 10^{10}$~G
to select DRPs is determined by
the maximum magnetic field observed for a recycled pulsar in a DNS:
$2.3 \times 10^{10}$~G for B1913+16~\citep{ht75a}. This cut-off
is valid, provided that higher magnetic field objects do not
evolve into either sample during their observational lifetimes.
Given the lack
of observational evidence for magnetic field decay in recycled
pulsars \citep{fk06}, there is no reason to suspect that this
cutoff will impose any significant bias into the {\it relative}
numbers of pulsars in each sample. 

An important selection effect for 
DRPs, however, is the ``contamination'' in the sample from the isolated
population of non-recycled pulsars \citep{kl00}. To quantify this effect, we
have used the results of recent studies of the normal pulsars
\citep{fk06,rl10} which predict the fraction of non-recycled
pulsars in the observed sample with $B<3 \times 10^{10}$~G to be
about 0.3\%. Given the present sample of $\sim 1500$ non-recycled
objects, we therefore expect 4--5 of these to have no binary
origin. We conclude that the best estimate for the observational
ratio of the DRP to DNS systems is therefore currently $r_{\rm obs} \sim 1$.

\subsection{Ages}

Our population synthesis of the DNS and DRP samples discussed below requires
some estimate of the likely radio lifetime of the mildly recycled pulsars
produced in these systems.
With their relatively weak magnetic fields, DNS and DRPs are expected to have
longer radio lifetimes by comparison to normal pulsars which are thought to
be a few 10s of Myr (e.g.~Lyne, Manchester \& Taylor 1985).
For all 20 objects listed in Table~1, the characteristic ages range between
100~Myr and 50~Gyr with mean and median values of 5.4 and 3.0~Gyr 
respectively. As recently discussed by Kiziltan \& Thorsett (2009). 
The characteristic
ages of recycled pulsars are likely to both significantly underestimate
and overestimate the true ages. The underestimate is caused by
secular accelerations which contribute to the observed
$\dot{P}$, while overestimates arise due to
sub-Eddington accretion in the progenitor phase
(Kiziltan \& Thorsett 2009) which result
in a birth period that is close to the current value. Taken as a whole,
the characteristic ages suggest a typical lifetime for the population
that is close to 10~Gyr, and we adopt this number in our
evolutionary simulations described in Section 3.3.

\subsection{Scale heights}

Despite the small-number statistics present in Table~\ref{tab:dnsdrp},
it is immediately apparent that the height above/below the Galactic
plane $z$ is, on average, significantly larger for DRPs than for DNS
binaries.  Taking the $z$ values from Table~\ref{tab:dnsdrp}, we find
$|z|= 200$~pc for the DNS systems compared to $|z|= 480$~pc for the
DRPs. This difference between the two populations could be explained
by a larger velocity dispersion for DRPs and/or longer radio
lifetimes.  We have already remarked that the radio lifetimes of DRPs
are likely to be longer than the DNSs. 
In the following section, we will also show on
evolutionary grounds that the expected velocity distributions of the
two populations are indeed fundamentally different.

\section{Binary population synthesis}\label{sec:simulation}

The population synthesis code we use, {\tt StarTrack}, was initially
developed to study double compact object mergers in the context of 
gamma-ray burst
progenitors (Belczynski, Bulik \& Rudak 2002b) and gravitational-wave
inspiral sources (Belczynski, Kalogera, \& Bulik 2002a). In recent years 
{\tt StarTrack} has undergone major updates
and revisions in the physical treatment of various binary evolution
phases, and especially the mass transfer phases.  The new version has
already been tested and calibrated against observations and detailed
binary mass transfer calculations (Belczynski et al.\ 2008a), and has
been used in various applications (e.g., Belczynski \& Taam 2004;
Belczynski et al.\ 2004; Belczynski, Bulik \& Ruiter 2005; Belczynski
et al. 2006; Belczynski et al.\ 2007). The physics updates that are
most important for compact object formation and evolution include: a
full numerical approach for the orbital evolution due to tidal
interactions, calibrated using high mass X-ray binaries and open
cluster observations, a detailed treatment of mass transfer episodes
fully calibrated against detailed calculations with a stellar
evolution code and updated stellar winds for massive stars (e.g., 
decreased mass loss from Wolf-Rayet stars that accounts for clumping; 
Hamann \& Koesterke 1998).

\subsection{Helium star evolution}

For helium star evolution, which is of crucial importance for the
formation of DNS binaries (e.g.~Ivanova et al.\ 2003; Dewi \& Pols
2003), we have applied a treatment matching closely the results of
detailed evolutionary calculations.  If the helium star fills its
Roche lobe, the systems are examined for the potential development of
a dynamical instability, in which case they are evolved through a
common envelope phase, otherwise a highly non-conservative mass
transfer ensues. We treat common envelope events using the energy
formalism (Webbink 1984), where the binding energy of the envelope is
determined from the set of He star models calculated with the detailed
evolutionary code by Ivanova et al.\ (2003).  For the case when the common
envelope is initiated by a star crossing the Hertzsprung gap, the
outcome of the common envelope is highly uncertain. Such stars do not
yet have well developed core-envelope structure and once the inspiral
process starts it may never end (whether there is enough of binary
orbital energy or not to eject the envelope) leading to the binary
component merger. If a merger is assumed, the evolution leads to a
very dramatic decrease of number of BH-BH binaries and a less
pronounced decrease for DNS systems (Belczynski et al.\ 2007). In this
study we allow either survival or we assume a merger in case the
Hertzsprung gap star is a donor in common envelope evolution. The
results are presented for both cases to test the influence of this
evolutionary uncertainty on recycled pulsar populations.

\subsection{Neutron star formation}

The full description of remnant mass calculation is given in Belczynski et
al. (2008; see their Sec. 2.3.1), and here we report only the most important
details.
Neutron stars form in a wide range of initial progenitor masses. For zero-age
main sequence (ZAMS) single
stars, neutron star formation begins at $M_{\rm ZAMS} \sim 7.5-8.5$ with
low mass neutron stars ($M_{\rm NS} \sim 1.2 \msun$) being formed via 
electron capture supernovae that involves core collapse of ONeMg core 
(e.g., Podsiadlowski et al. 2004). For higher
initial masses, neutron stars form through core collapse of FeNi core; for  
$M_{\rm ZAMS} \sim 8.5-18 \msun$ neutron stars form with mass $M_{\rm NS}
\sim 1.3 \msun$, while for heavier progenitors, $M_{\rm ZAMS} \sim 18-20 
\msun$, neutron stars form with $M_{\rm NS} \sim 1.8 \msun$. Such a bimodal
distribution is explained by the different element burning in the cores of 
massive stars that results in the formation of lower mass FeNi cores for 
lighter stars where the central temperature is not high enough for more 
effective burning (Timmes, Woosley \& Weaver 1996). Although the
majority of neutron stars with mass determinations
fall in the range $\sim 1.2-1.4 \msun$ there are a
number of pulsars for which higher masses are likely
exist (e.g., $\sim 1.9 \msun$,
Vela X-1, Barziv et al. 2001; $\sim 1.7-1.8 \msun$ for Terzan~5 I and J,
Ransom et al. 2005 ), although the error bars are still large (e.g., Lorimer 
2008). For progenitors with $M_{\rm ZAMS} \gtrsim 20 \msun$, the fallback of
material (e.g., Fryer \& Kalogera 2001) during supernova explosion may
increase a proto neutron star mass such that it collapses to a black hole. 
We assume the maximum neutron star mass to be $M_{\rm NS,max} = 2.5 \msun$,
and then progenitors with masses $M_{\rm ZAMS} \gtrsim 21 \msun$ form black
holes. 

During formation, a neutron star receives a natal kick that along with
the mass loss from the exploding star may lead to disruption of a binary
system, if an exploding star is a binary component. Since natal kicks are a 
major factor in the disruption of binaries, we can also expect them to play a crucial
role in determining the DNS/DRP ratio. To investigate this issue in detail, we 
use our standard evolutionary model and vary the spread of the underlying
natal kick distribution. In a recent analysis of the pulsar birth velocity
distribution, Hobbs et al. (2005) found that the observed sample
could be well described by a single Maxwellian with 
$\sigma_{\rm Hobbs} = 265$ km s$^{-1}$. It is not clear whether this
distribution may be directly applied for stars in binaries since the observed 
pulsars are single and we do not know how many have originated from binaries. 
Also, if a given single pulsar originates from a binary, mass loss and orbital 
velocity at the time of the supernova explosion disrupting a binary will factor into 
the final pulsar velocity (in addition to the natal kick it has received). 
Therefore, we employ the observed distribution just in one of our calculations 
(model A) assuming that all neutron stars that form in FeNi (regular) core collapse 
supernovae receive a natal kick drawn from a distribution with $\sigma_{\rm CC}= 
\sigma_{\rm Hobbs}$, and then we decrease the kicks for the sequence of
models:
$\sigma_{\rm CC} = 0.75 \sigma_{\rm Hobbs} = 199$ km s$^{-1}$ (model B), 
$\sigma_{\rm CC} = 0.5  \sigma_{\rm Hobbs} = 133$ km s$^{-1}$ (model C),
$\sigma_{\rm CC} = 0.25 \sigma_{\rm Hobbs} = 66$ km s$^{-1}$ (model D), 
and $\sigma_{\rm CC} = 0$ km s$^{-1}$ (model E). 
In each of the above models, we assume that there is no natal kick in the case 
of neutron star formation through electron capture supernova ($\sigma_{\rm
ECS}=0$) as recent numerical simulations indicate that explosion energy 
may be much smaller in such a case (e.g., Dessart et al. 2006; Kitaura, Janka
\& Hillebrandt 2006). 

\subsection{Mass accretion}

In the evolutionary scenarios for DNS and DRP progenitor binaries we consider 
the amount of mass accretion onto the neutron star to be relatively modest. This 
naturally follows from the fact that a first born neutron star always has a 
(much) more massive
companion. In the event of RLOF, most often it proceeds on a thermal 
timescale of the massive donor and with super-Eddington mass transfer rates
and only a small fraction of the transferred material ($\sim 1\%$) is
usually accreted onto a neutron star. Even in case of mass transfer on a
nuclear timescale of the donor, the duration of RLOF is usually so short that
not much is accumulated on a neutron star (short lifetime of massive donor). 
Additionally, in the case of dynamically unstable events that lead to common
envelope evolution it was pointed out (e.g., Ruffert 1999; Ricker \& Taam 
2008) that only a small amount of mass may be accreted onto a compact object 
(a black hole or a neutron star). We calculate a Bondi-Hoyle accretion rate
onto a compact object during a specific common envelope event (Belczynski et
al. 2002a) and then allow for accretion at the level of only $10\%$ of the 
calculated rate.  
Such an approach leads to the formation of DNS population with a neutron star
of low mass $1.2-1.4 \msun$ that reproduces rather well the observed systems 
(e.g., Belczynski et al. 2008b).
First born neutron stars in DNS and DRP populations usually accrete (if any
RLOF/common envelope was encountered) $\sim 0.01-0.1 \msun$. Although an 
exact amount of mass to recycle a pulsar is not well established (e.g., 
Zdunik, Haensel \& Gourgoulhon 2002; Jacoby et al. 2005), we assume that 
if a neutron star accreted over $0.05 \msun$ it becomes a mildly recycled
pulsar with a lifetime of 10~Gyr as discussed above.

\subsection{Simulation specifics}

In each calculation we evolve $2 \times 10^6$ massive binaries
($6<M_1<150 \msun$, $4<M_2<150 \msun$), in which the primary mass ($M_1$)
is drawn from power-law initial mass function with slope $-2.7$, and
the secondary is chosen via a flat mass ratio distribution
($q=M_2/M_1$). All binaries are allowed to be initially eccentric ($f(e) 
\propto 2e$), while their separations are drawn from a flat distribution
in logarithm (i.e., $\propto 1/a$) and reaching maximum of $10^5
R_\odot$. All stars are evolved with solar-like metallicity ($Z=0.02$)
and are assumed to form in the Galactic disk (i.e., continuous star
formation through the last 10 Gyr). We perform a time cut at the present
time and count the numbers of DNS and DRP hosting an active recycled
pulsar. The numbers presented throughout our study are not
calibrated as we are mostly concerned with the ratio of DNS to
DRP. However, very easy calibration may be performed on these numbers
to represent the entire synthetic population of active recycled
pulsars in Galaxy. The presented numbers need to be multiplied by
a factor of $\sim 40$ to give the star formation rate observed currently
in the disk of the Galaxy ($\sim 3.5 \msun$ yr$^{-1}$) or result in
supernova II and Ib/c rate estimated for a Milky Way-type Galaxy
($\sim 0.02$ yr${-1}$).

We consider only the formation of recycled pulsars in massive star
populations, i.e., stars that can form neutron stars/black holes.
Recycled pulsars can also be formed in binaries with a companion star
that is not massive enough to form a second neutron star/black hole,
e.g., neutron star low- or intermediate-mass main sequence star,
neutron star low mass-mass helium star or neutron star white dwarf
binaries. However, any of these binaries cannot form a single recycled
pulsar (or one in a DNS), unless some rather exotic scenarios are
considered (e.g., evaporation of a white dwarf by a neutron star).

In the first scenario,
two stars of similar mass ($\sim 10-20 \msun$) start the evolution. 
The more massive primary initiates the first (stable) RLOF episode, 
potentially rejuvenating the secondary before forming the first neutron star.
Very often this is a low-mass neutron star ($\sim 1.2 \msun$) formed 
through electron capture supernova. The secondary then initiates a second
RLOF episode. This time, due to high mass ratio ($q \gtrsim 5$; the ratio
of the secondary star and neutron star masses), the common envelope phase
is initiated. The system emerges as a close neutron star--helium star
binary. The neutron star has accreted some material while moving through
the envelope of the secondary ($\sim 0.05 \msun$). The helium star
expands and initiates the third RLOF. This time it may be either
a dynamically stable or unstable event. In the case of the common
envelope there is a large uncertainty whether such a system survives or
not. Frequently the helium donor is crossing the helium Hertzsprung gap and
has not yet developed a clear core-envelope structure that is
needed to halt an inspiral during this phase (for details see
Belczynski et al.\ 2007).

In the above example we have presented our most efficient scenario for 
the DNS/DRP formation. However, in our population synthesis calculations we
include a number of various formation channels (for example see Table 3 
of Belczynski et al. 2002a). In particular, we include the ``Brown'' 
channel (Brown 1995) of the DNS formation of two almost equal mass stars
that avoids common envelope with a NS accretor (e.g., see channel NSNS:09 
of Belczynski et al. 2002a). This channel was also followed by other groups 
(e.g. Dewi, Podsiadlowski \& Sena 2006) and it was found that the rates 
are generally smaller than predicted in the original work. This study
predicts much lower rates for this particular channel as compared with
the original Brown work. The difference stems from the fact that the early  
estimates on the amount of accretion ($\sim 1 \msun$; Bethe \& Brown 
1998) onto NS in the common envelope phase were most likely overestimated.  
With high accretion rates, in all the classical channels (like the 
one we presented above) that involve a common envelope phase, a NS accreted 
enough to collapse to a BH avoiding the NS-NS formation. As discussed in 
section 3.3 we follow recent estimates of accretion in common envelope ($\sim 
0.1 \msun$; Ruffert 1999; Ricker \& Taam 2008) and allow for the NS-NS 
formation along variety of channels. 

In our modeling, and in particular in the presented example of the DNS/DRP  
formation, the first NS forms predominantly in electron-capture supernova 
while the second NS is formed either in regular core-collapse or electron 
capture supernova. This is consistent with the original ideas of Pfahl et 
al. (2002a) who argued that neutron stars formed in some specific high-mass
X-ray binaries are formed with a low kick. The DNS/DRP progenitors evolve
through a high-mass binary phase after the first NS formation. The potential 
explanation was discussed by Podsiadlowski et al. (2004), who pointed out
that depending on initial binary orbital period, the first star may either
form a NS via electron capture or regular (FeNi) core-collapse supernova. 
Our result stems 
from the fact, that majority of the DNS/DRP progenitors are still found on
relatively wide orbits at the first SN explosion and if any significant kick
is imparted on a NS a given system is most likely disrupted, barring the 
formation of either DNS or DRP. Hence, the systems that are in the mass 
range of electron-capture NS formation at the time of the first supernova 
are naturally selected in the DNS/DRP formation. At the time of the second 
supernova the progenitors are usually very close binaries, and the effect 
of kicks is not as severe on the system survival as during the first SN. 
In our simulations it is found that the second SN is dominated by regular
core-collapse with smaller contribution of the electron-capture NS formation. 
Some known DNS have significant eccentricities (B1913+16: $e=0.617$;
J1811-1736: $e=0.828$) that are indicative of a significant natal kick at 
the second SN. Even for some moderate eccentricity systems (e.g. B1534+12: 
e=0.274) high natal kicks are derived (e.g., Stairs et al. 2006).  
For some low-eccentricity systems (e.g., J0737-3039: $e=0.088$) the low
kicks are claimed (e.g., Piran \& Shaviv 2005), but high kicks at the 
second SN can not yet be excluded (Willems et al. 2006).  
If it will turn out that the second supernova in DNS binary progenitors is
predominantly electron-capture SN with a low (or no) kick, it will allow us 
to put strong constrains on the initial mass range (broader than assumed in
this study) for this mode of NS formation.

\subsection{Results}

The results of two models are presented. In the first model,
we allow for such a survival. In the second model, such systems are assumed to 
merge. During the third RLOF episode the neutron star accretes some more 
material from its companion ($\sim 0.05 \msun$). The first neutron star, 
which accreted about $\sim 0.1 \msun$, has most probably become a recycled
pulsar. After (or during) the third RLOF the companion star explodes
forming the second (non-recycled) neutron star. The second neutron star
is formed in regular supernova explosion/core collapse. Regular core collapse
supernovae (stars forming FeNi cores) are more massive than stars
forming neutron stars through electron capture supernovae
(semi-degenerate ONeMg cores). Early on in the evolution of a
progenitor (first RLOF) there is a mass ratio reversal, and in fact it
is expected that the secondary is in the end the more massive star. After
the third RLOF, the system becomes very close and many such systems have a
good chance of supernova explosion survival. The systems that are disrupted
at the second supernova produce two single neutron stars, while surviving
systems form DNS binaries. Depending upon the amount of accretion onto
a first-born neutron star either a DRP or a DNS recycled pulsar may form.

In the second scenario, the evolutionary history is almost the same as
presented in scenario 1 with the difference being that the secondary star forms
a black hole.  The stars are initially more massive, and then during
the first RLOF episode the secondary accretes enough mass to form a black
hole at the end of its evolution. As before, the primary forms a neutron
star and it is the first formed compact object in a system.  Such a
scenario may lead to the formation of either a NS-BH binary or, if a
system is disrupted upon black hole formation, a single recycled pulsar.  
This scenario is rather inefficient ($\sim 0.2\%$) as
compared to scenario 1 ($\sim 98.8\%$) in the formation of DRPs. 
This is due to the fact that scenario 2 is allowed
for only very specific combinations
of progenitor masses, i.e. both component masses need to be very
close to the mass limiting neutron star and black hole formation.

\subsubsection{DNS/DRP numbers}

In Table~2 we list the intrinsic (i.e.~with no detection biases accounted
for) numbers of DNS with recycled pulsar, DRP, and their ratio as
obtained in the population synthesis calculations.  Numbers are listed
for all our models (natal kick velocity varied), and within each model
we give a range corresponding to common envelope uncertainty; the high
numbers of DRP/DNS correspond to calculations in which survival
through the common envelope is allowed for Hertzsprung gap donors, while
the low numbers correspond to the assumption of a merger during such a
phase.

The predicted number of DNSs increases with decreasing natal kicks, as
the kicks are very effective in disrupting potential DNS
progenitors. For model A (high kicks) most of the potential DNS
progenitors are disrupted in the first supernova explosion ($98\%$), while
a much smaller fraction are disrupted at the second supernova
($1.8\%$), and only a very small fraction ($0.2\%$) survive and form
DNSs.  For model C (intermediate kicks) the disruption is $97.6\%$, and
$1.9\%$ in first and second supernova, and $0.5\%$ of the systems survive to
form DNSs.  The very high disruption rate at the first supernova comes from
the fact that many binaries at that time still have rather wide
orbits. At the second supernova, not only the wide binaries were
eliminated by first supernova disruptions, but also consecutive RLOF
and/or common envelope phases decrease the separation between binary
components.  For model E (no natal kicks) in which disruptions are due
only to mass loss during supernova explosion, we find $16.3\%$, and
$79.5\%$ disruptions occurring in the first and second supernova, and
$4.2\%$ systems survive as DNS binaries.  From the virial theorem,
disruption by mass loss alone requires about $50\%$ of the total binary
mass to be lost. It is therefore much easier to disrupt binaries at
the second supernova during which time a first born neutron star is a
less massive binary component.  For the DNS population, most of the first
born neutron stars ($\gtrsim 80\%$) accreted enough mass to host a
recycled pulsar, and only these systems are listed in Table~2.

The number of DRPs at first decreases with decreasing kick velocity as
the smaller kicks are less effective in releasing recycled pulsars
from progenitor binaries.  However, for very low kicks ($\sigma < 100$
km s$^{-1}$) the number of DRPs increases with decreasing kick
velocity. This is a natural effect of the higher disruption rate
during the second supernova (high release of recycled neutron stars
from binaries) relative to the first supernova for low or zero kicks as
explained above. Progenitors of disrupted binaries are on average
wider (easier to disrupt) than progenitors of a DNS. Since they are
wider, the stars interact less (less mass transfer) and in the end not
so many first born neutron stars are recycled. If we consider just
progenitors that are disrupted at the second supernova (so the ones that
have a chance to produce a solitary recycled pulsar) we find that only
$\sim 1-20\%$ of the disrupted binaries form a DRP. Smaller fractions
are found for models with no or low natal kicks as wider
non-interacting systems more readily survive the first supernova.

\subsubsection{DNS/DRP spatial velocities}

\begin{figure}
\psfig{file=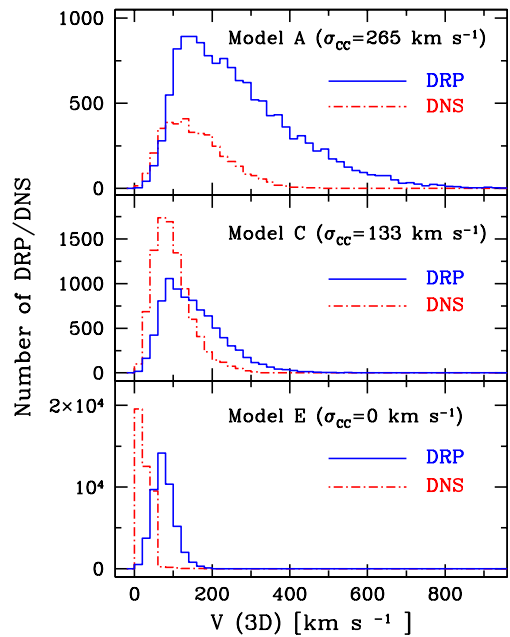,width=9cm,angle=0}
\caption{Predicted velocity distributions for DRP and DNS populations for
different models. The presented distributions correspond to evolutionary
calculations in which common envelope phase with Hertzsprung gap donor is
allowed (the shape of distributions is virtually the same for calculations
in which such phase is assumed to lead always to a merger).
}
\end{figure}

In Fig.~2 we present the resulting velocity distributions for DNS and DRP
populations.  Results for models with high (model A), intermediate (C)
and zero (E) natal kicks are presented. The presented distributions
are obtained for evolutionary models that allow for the survival
through the common envelope phase with Hertzsprung gap donor (i.e.,
they correspond to the higher numbers in Table~2).  The distributions
for alternative treatment of such a phase are qualitatively very
similar. The velocities we present are those that the DNSs and DRPs
obtain during the two supernova explosions from mass loss and/or natal
kicks. In other words they can be understood as an extra velocity
component that should be added to a typical Galactic velocity for a
given object.

As can be seen from Fig.~2,
that velocities of DRPs are higher than the velocities
gained by DNSs. Also the distributions are broader for DRPs. These
general features hold for all considered models. Progenitors of DRPs
are disrupted at the second supernova explosion mostly due to a rather
high and/or unfavorably (e.g., perpendicular to orbital plane: high
disruption probability) placed kick. Additionally, mass loss from an
exploding star and its orbital speed at the time of explosion factor
into the final DRP velocity (the full description of the velocity
calculation is given in Belczynski et al. 2008; see their Section
6.3).  The progenitors of a DNS survive two supernova explosions and
tend to receive smaller or favorably (for survival) placed kicks and
their final velocities are on average smaller than these of DRPs.

We can see that the natal kicks play the major role in setting the
spatial velocities of both populations. If no natal kicks are applied
at supernova explosions (model E) we see that the mean velocities are
rather small: 26 and 76 km s$^{-1}$ for DNS and DRP populations,
respectively. However, if even intermediate kicks are applied (model
C) we note a significant increase of the mean velocity: 99 and 154 km
s$^{-1}$ for DNS and DRP populations, respectively.

The substantial differences in the DNS and DRP velocity distributions
may lead to a different detection probability for both populations. If
there is any observational bias against detecting either DNSs or DRPs
we need to take it into account and revise our intrinsic ratio $r_{\rm int}$
before attempting a comparison with the observed ratio $r_{\rm obs}$ below.

\subsubsection{Orbital period distributions of DNS binaries}

\begin{figure}
\psfig{file=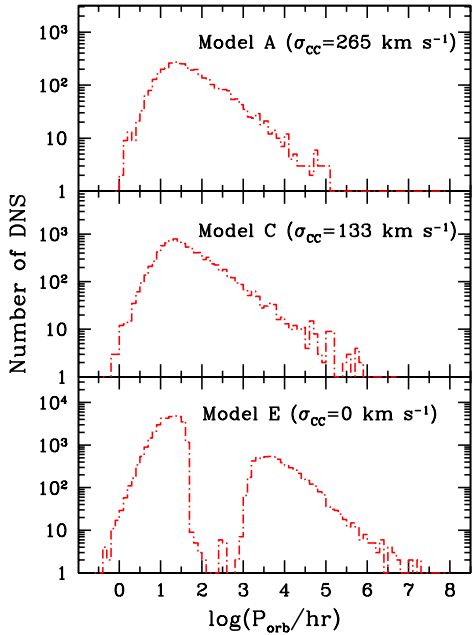,width=9cm,angle=0}
\caption{Predicted period distribution for DNS binaries hosting a recycled
pulsar for the same models as in Fig.~2. 
}
\end{figure}

In Fig.~3 we present the orbital period distributions for the DNS
population (only binaries hosting a recycled pulsar are shown) for
models A, C, E. For all of the models, the orbital
periods extend to high values (see mean and standard deviations for
the distributions), although the majority of systems are found at
small periods (medians are in the range $\sim$~10--25~hr). 
In the following text we explain the shape of the orbital
period distributions for DNS
for various natal kick models. In particular, it is interesting that models
that involve natal kicks (A-D) have continuous distributions, while model
E which has no natal kicks shows a 
bimodal bimodal period distribution (see Fig.~3). 

Prior to the second SN explosion (i.e.~just before DNS formation), the period 
distribution  is bimodal for all models. Long-period systems ($P_{\rm orb} 
\sim 100$ hr) originate from progenitors evolving via classical DNS formation
channels (e.g., Bhattacharya \& van den Heuvel 1991) that involve only two
RLOF episodes, while short-period systems ($P_{\rm orb} \sim 10$ hr) 
originate from new DNS formation scenarios (Belczynski \& Kalogera
2001; Ivanova et al. 2003; Dewi \& Pols 2003) that involve three RLOF
phases. The extra RLOF phase (that is the last phase before the second SN
and the DNS formation) tends to make orbital separation (and orbital period) 
decrease and thus creating a bimodal period distribution prior the second
SN. There is also an additional effect; for the binaries evolving through the
extra RLOF the donor star that is just about to explode in SN and form the 
second NS is of a very low mass (mass loss in the RLOF). At the time of
the second SN explosion, the exploding star loses on average $\sim 0.5
\msun$ in the new formation scenarios, while it loses $\sim 2.5 \msun$ for 
classical formation channels. 

The bimodality, in orbital period and mass ejection, at the second SN leads 
to the bimodal distribution of DNS orbital periods for model with no natal
kicks (E). The systems with short pre-SN orbital periods have small mass 
ejection and thus they tend to survive the SN with not much change of 
orbital parameters and they end up as close DNS with $P_{\rm orb} \lapp 
100$ hr (the new DNS).  On the contrary, the binaries with the long pre-SN 
periods have significant mass ejection and thus the orbit increases and 
gains high eccentricity at the second SN and these binaries form rather 
long-period DNS binaries ($P_{\rm orb} \gtrsim 1000$~hr).
  
For models with natal kicks (A, B, C, D), the progenitor binaries follow 
similar evolutionary channels (the classical and new formation scenarios).   
At the time of the second SN there is similar mass ejection, however the 
additional natal kicks (with various directions and magnitudes) tend to
smear the two peaks in pre-second SN period distribution out. 
The distributions peak at $\sim 10$ hr (preserving the shape of pre-SN period
distribution for the closest binaries; the hardest to disrupt and change the
orbit) and then it falls off with the increasing period. Additionally to the
smearing, the effect of increasing natal kicks can be observed with 
distributions terminating at shorter periods and general decrease of DNS with 
increasing kicks (enhanced SN binary disruptions).

\section{Observational selection effects}\label{sec:selfx}

The evolutionary 
models described above do not take into account the observational
selection biases which are known to be significant for the
radio pulsar population \nocite{lbdh93} (see, e.g., Lorimer et al.\ 1993). 
In this section, we investigate the various factors which might
affect the relative populations of DRPs and DNS systems. 

\subsection{Basic assumptions}

We begin
by making an important assertion: the spin-down, luminosity and
beaming evolution are the same for both DRPs and DNS binaries.
This simplification should hold for any model in which the two populations
follow an accretion phase that determines the initial spin period
of the recycled pulsar. The subsequent spin-down evolution is
the same regardless as to whether the binary system was disrupted
at the end of mass transfer or not. The spin period, radio luminosity
and beaming fraction of the recycled pulsar should be identical.

Given this premise, there are only two possible differences that
affect the detectability of DRPs as opposed to DNS binaries. Firstly,
Doppler smearing in the binary orbit during a survey integration time
will significantly degrade the sensitivity to DNS binaries with the
shortest orbital periods. Secondly, the different predicted velocity
distributions for DRPs and DNS systems will result in a larger
distance from Earth for the DRPs and hence smaller average flux
density than the DNS binaries. These two effects act in opposite
ways. Doppler smearing tends to select against the detection of DNS
binaries relative to DRPs, while the higher velocities of DRPs
compared to DNS systems selects against DRPs.

\subsection{Monte Carlo simulation}

To quantify the strength of these effects on
the observed sample, we carried out a simple Monte Carlo simulation
described in detail below which compares the relative
detectability of DNS binaries to DRPs in radio pulsar surveys.
The ratio of the numbers of detectable DNS binaries to DRPs
then provides us with an estimate of the correction factor
$f_{\rm obs}$ required to scale the intrinsic DNS/DRP ratios listed
in Table 2 and compare them with the observed DNS/DRP ratio
as defined in Equation 1.

The Monte Carlo code we use for this analysis,
{\tt psrpop}, is a freely available software package to model 
pulsar populations and radio survey selection effects \citep{lfl+06}.
We have recently extended the scope of {\tt psrpop} to model the
kinematical and spin-down evolution of pulsars to investigate
a number of issues in pulsar statistics \citep{rl10}.
For the purposes of this study, we generated two populations 
with identical numbers of
recycled pulsars. For both populations, we assumed a dipolar spin-down
evolution from an initial spin period of 20~ms for a magnetic
field strength of $2 \times 10^{10}$~G. The radio luminosity of
each pulsar at 1.4~GHz was assumed to be $10^3$~mJy~kpc$^2$. Each
model pulsar was then allowed to move in a model of the Galactic
gravitational potential for a random age of up to 1~Gyr, the typical
radio lifetime of a recycled pulsar. The exact details of these
assumptions have no effect on our results, since we are only 
concerned with the relative numbers of each population that are
detectable and we are assuming that the spin-down, luminosity and
beaming evolution are identical for each population. 

To model the Doppler smearing due to binary motion for the DNS systems, we 
follow the approach of \citet{jk91}. In this framework, for a given survey
integration time and set of orbital parameters, a signal-to-noise 
reduction factor $\gamma$ is computed by comparing the response in the Fourier
domain between an isolated pulsar and a binary system. We assumed a 
Fourier spectrum which is optimally summed for 16 harmonics (which is typical
for pulsar search detections), and averaged over all orbital phases
and inclinations assuming circular orbits with randomly inclined planes
with respect to the line of sight. The circular orbit assumption is
made for computational convenience and is an excellent approximation for
DNS systems like J0737$-$3039 and J1906$+$0746. For the more
eccentric systems such as B1936+16 and J1756$-$2251, this approach
provides a more approximate but conservative measure of $\gamma$. We 
defer a full extension of Johnston \& Kulkarni's analysis for
elliptical orbits to a future study.

As described in \citet{jk91}, the factor $\gamma$ can be computed
for surveys with and without coherent acceleration searches. Most of
the surveys considered below had relatively short integration times
and did not adopt acceleration searches. However, the Doppler smearing
effects can be significant for short orbital periods and this is
taken into account in our simulations. For the Parkes multibeam pulsar
survey of the Galactic plane which had 35-minute dwell times \citep{mlc+01},
acceleration search techniques were applied \citep{fsk+04}. As discussed
by Faulkner et al., the ``stack search'' method used in this analysis
is typically 25\% less efficient than a fully coherent acceleration search.
We take this factor into account when computing $\gamma$ for this
survey. The two other surveys with fully coherent acceleration searches
we consider are the ongoing Pulsar Arecibo L-band Feed Array (PALFA)
survey \citep{lsf+06} and the Green Bank 350-MHz drift scan survey
\citep[GBTDRIFT;][]{asr+09}.

\subsection{Results}

With the above set of assumptions, we considered two populations.  For
one population (model DNS) we drew velocities from the distribution
predicted for the DNS binary systems produced in model A (see
Fig.~2). The orbital period distribution we used for these systems was
an analytic form of the predicted distribution of DNS orbital periods
from model A shown in Fig.~3. For this model, we find that the
cumulative number of systems as a function of orbital period is well
approximated by the simple function $N(<P_b)=P_b/(1+P_b)$, where $P_b$
is the orbital period in days.  For the second population (model DRP),
we drew velocities from the distribution predicted for the DRPs from
the same model.  For both populations, we computed the final position
and expected flux density of each model pulsar and used models of a
variety of recent pulsar surveys to calculate the number of detectable
pulsars.

\begin{table}
\caption{\label{tab:simulations}
Results of simulation runs for the model DRP and DNS populations.
From left to right, the columns give the survey name, integration
time ($t_{\rm int}$), whether acceleration searches were used (AC), simulated
number of detected DNS binaries ($N_{\rm DNS}$), number
of detected DRPs ($N_{\rm DRP}$), and the correction factor
$f_{\rm obs} = N_{\rm DNS}/N_{\rm DRP}$.
The surveys considered are: an Arecibo drift-scan survey (AODRIFT),
the Green Bank Telescope drift-scan survey (GBTDRIFT), the Parkes
Multibeam surveys of the Galactic plane (PMSURV) and high latitudes (PHSURV),
the 70-cm Parkes Southern Sky Survey (PKS70), the Pulsar Arecibo
L-band Feed Array survey (PALFA) and the Swinburne Parkes Multibeam mid
and high latitude surveys (SMMID and SMHI).
}
\label{tb:simulations}
\begin{tabular}{llllll}
\hline
Survey & 
$t_{\rm int}$ (s) &
AC &
$N_{\rm DNS}$ & 
$N_{\rm DRP}$ & 
$f_{\rm obs}$ \\
\hline
AODRIFT  & 45 & No & 13377 & 11559 & 1.16 \\
GBTDRIFT & 180& Yes&  9553 & 8139 & 1.17 \\
PMSURV   & 2100&Yes& 26715 & 16435 & 1.63 \\
PHSURV   & 524 & No&2227 & 2141 & 1.04 \\
PKS70    & 180 & No&35783 & 32613 & 1.10 \\
PALFA    & 278 &Yes&4101 & 2391 & 1.72 \\
SMMID    & 524 & No&21641 & 17160 & 1.26 \\
SMHI     & 524 & No&9781 & 10364 & 0.94 \\
\hline
\end{tabular}
\end{table}

Our results are summarized in Table \ref{tab:simulations}
 where we calculate the number of
each model pulsar population detected ($N_{\rm DNS}$ and $N_{\rm
DRP}$) and the ratio of the two populations $f_{\rm obs} = N_{\rm DNS}/N_{\rm
DRP}$. The absolute values of $N_{\rm DNS}$ and $N_{\rm DRP}$ are, of
course, arbitrarily high and chosen to be so to minimize statistical
fluctuations. Given our assertion of that the spindown and beaming of
the two populations should be identical, it is their ratio that is
of astrophysical significance.  As can be seen, depending on the
survey parameters, there is a variation in the ratio in the range
0.9--1.7. Surveys along the Galactic plane (i.e.~PMSURV and PALFA)
show the largest bias in favor of detecting DNS systems over DRPs
(i.e.~the largest values of $f_{\rm obs}$), while surveys away from the plane
(e.g.~PHSURV and PKS70) show much less of a bias. Indeed, for the
Swinburne high-latitude survey \cite{jbo+07}, the bias is slightly
tipped in favor of detecting DRPs. Taken as a whole, the
average value of $f_{\rm obs}=1.25$ in Table \ref{tab:simulations}
suggests that we might expect
to detect around 25\% more DNS binaries than DRPs due to selection effects
alone. Thus the observational bias is a relatively small perturbation
on top of the intrinsic ratios found in the population syntheses
listed in Table 2.

Although we have only explicitly considered the velocity and
orbital period distributions from model A in this study, similar
results are found when the other model parameters are used as
input. In summary, while we find that observational biases exist which
favor the detection of DNSs over DRPs, their magnitude is
not sufficient to significantly skew any underlying ratios
of the two populations. 

This allows us to revisit the results from the population syntheses 
presented in Table 2. As discussed earlier the observed ratio is 
$r_{\rm obs} \sim 1$. If we fold this with observational bias factor 
($f_{\rm obs} = 1.25$) the intrinsic ratio of DNS to DRPs is of the 
order $r_{\rm int} \sim 0.8$ (DRPs slightly dominate over DNS). 
The comparison with Table 2, that shows the predicted ratios $r$, 
shows two things. First, the model with natal kicks adopted from 
observations of single pulsars (model A) does not reproduce the
intrinsic ratio ($r=0.3$). Second, the models that are close to the
intrinsic ratio (model B and C, $r=0.5,\ 1.1$ respectively) indicate 
that the 1--D dispersion of kick velocity distribution is of the 
order of $\sigma_{\rm CC} \sim 170$ km s$^{-1}$ (approximately the 
mid point between model B and C).

\section{Conclusions}\label{sec:conclusions} 

We have investigated the statistics of disrupted recycled pulsars
(DRPs) and double neutron star (DNS) binaries where it is observed
(see Section 2) that there are comparable numbers of DNS binary
systems and DRPs.  From a population synthesis of neutron star
formation and evolution in binary systems (see Section 3), regardless
of the assumed natal kick velocity distribution for neutron stars, we
find that the velocity dispersion of DRPs is significantly higher than
for DNS binaries.  Using the resulting orbital period and velocity
distributions in a model for radio pulsar selection effects (see
Section 4), we found that there is a small ($\sim 25\%$) observational
bias in favor of detecting DNS binaries over DRPs.
The difference in velocity dispersions for the two
populations lead to a bias towards observing DNS binaries
which are, on average, closer to the Earth than DRPs. This
conclusion holds even after taking into account that DNS
binaries are harder to detect due to Doppler smearing.

Based on these results, we conclude that our models in which the natal
kick velocity dispersion above 200~km~s$^{-1}$ (e.g.~model A) or below
100~km~s$^{-1}$ (e.g.~model D) are inconsistent with the observations.
Model A predicts significantly more DRPs than are observed, while
model D predicts significantly more DNS binaries than are
observed. Models with intermediate natal kicks (e.g.~models B and C)
provide a better match to the observed sample. The typical kicks 
(from regular (FeNi) core collapse supernovae) that match the 
observed intrinsic ratio are 
of the order of $\sigma_{\rm CC} \sim 170$ km s$^{-1}$.
In all our simulations we have included the formation of neutron stars 
through electron-capture supernovae with no natal kicks. The initial mass 
formation range for electron-capture NS formation was adopted from 
Hurley et al. (2000) and Eldridge \& Tout (2004a,b) for single stars 
($M_{\rm zams}=7.6-8.3 \msun$; see Belczynski et al. 2008), and it is 
naturally extended by binary evolution. If the adopted range for
electron-capture supernovae was much broader than adopted here, it would 
be possible to explain the the observed intrinsic ratio of DNS to DRPs 
with much higher regular (FeNi) core-collapse supernova kicks. Independent of details 
of the electron-capture NS formation, our results strongly indicate that 
NS natal kicks (whether formed via regular core collapse or electron-capture) 
are much lower in binaries than for single pulsar population. 

The research into natal kicks and the formation of single recycled pulsars 
by supernova disruption has already a long history (e.g., Bailes 1989). 
The most recent study by Dewi et al. (2006) is not directly comparable to 
our results. Dewi et al. (2006) have studied only one (alternative) scenario
of the DNS formation and based their conclusions on the information available 
at the time of their study (7 DNS and 2 DRP). However, it is interesting to
note that the results from Dewi et al. (2006) 
show slightly higher DNS birth rates as
compared to disruption rates (DRP formation) for example for the model in 
which natal kicks are drawn from distribution with $\sigma=190$ km s$^{-1}$
(see the first entry in their Table 1). This result is fully consistent with
the current observational dataset, presented here, and it is similar to our
finding for the two models B and C that also match the observations.

Our conclusion that small neutron star kicks are required to explain
the DNS--DRP populations adds to the growing body of evidence that
such kicks are required in the formation of close binaries. The original
impetus for this idea was provided by Pfahl et al.~(2002a) based on
statistics of high-mass X-ray binaries with orbital periods longer
than 30 days. A similar requirement was seen for the neutron star
population in globular clusters (Pfahl et al.~2002b) in  which the low
escape velocities of the clusters predict many fewer pulsars than
are observed (Drukier 1996).
The notion of electron-capture supernovae, which naturally
produce such neutron stars with lower velocity kicks was subsequently
introduced by Podsiadlowski et al.~(2004). More recent evidence
for low-velocity kicks in binary systems has been inferred from
the spin-eccentricity correlation seen in DNS binaries (Faulkner
et al.~2005; Dewi et al.~2005) and in a starburst activity of
high-mass X-ray binaries in the Small Magellanic Cloud
(Linden et al.~2009).

The process of the formation of neutron stars via electron capture 
supernovae is highly uncertain; the magnitude of kicks (if any; e.g., 
Dessart et al. 2006; Kitaura et al. 2006), the initial star mass range 
for these explosions 
\citep[e.g.,][]{nom87} or physical conditions like rotation 
\citep[e.g.,][]{prps02} at which such a process may occur. We have included 
the possibility of formation of neutron stars via this process in our 
calculations and we have assumed that these types of explosions are not 
connected with any significant natal kicks. Still, our results indicated 
that there are too many DNS progenitor binary disruptions to reproduce 
the approximately equal observed numbers of DNSs and DRPs. The 
disruptions are mostly due to the natal kicks received by neutron stars in 
regular (FeNi) core collapse supernovae. It naturally led us to conclude 
that the regular core 
collapse natal kicks are smaller than it is usually believed in the case of 
close binaries. However, it needs to be noted that a similar effect (fewer 
disruptions) may be achieved via extended formation of neutron stars via 
electron capture supernova with negligible (or no) kicks (e.g. Podsiadlowski 
et al. 2004). Whether it is rather small regular core collapse kicks, or an excess 
of formation of neutron stars via electron capture supernovae or some other 
process, it seems to be clear that the kicks that operate in close 
interacting binaries (like for progenitors of DNS) are significantly smaller 
than the ones inferred for the population of Galactic single pulsars (e.g., 
Hobbs et al. 2004). It is interesting to note that the supernova
hydrodynamical simulations predict increase of asymmetries with rotation
(Chris Fryer, private communication). If natal kicks are connected to 
assymetries in regular (FeNi) core collapse then NS in binaries (fast rotation) are 
expected to receive larger kicks then single stars, opposite to what seems 
to be inferred from observations. On the other note, the kick mechanism 
may be totally different for an exploding $\sim 10-15 \msun$ H-rich star 
(single NS progenitor) and a $\sim 2-3 \msun$ He-rich star (binary). 

\section*{Acknowledgments}
We would like to thank T.Bulik, W.Kluzniak and P.Haensel for useful discussions 
on pulsar populations. 
This research was partially supported by a WVEPSCoR Research Challenge Grant
held by the WVU Center for Astrophysics.
KB acknowledges the partial support from the Polish MSHE grants N N203 302835 
and N N203 511238 and by LANL under contract No. DE-AC52-06NA25396.
The pulsar parameters used in Table 1 were obtained from the ATNF Pulsar 
Catalogue~\citep{mhth05}.

\end{document}